\documentstyle[epsfig]{l-aa}
\input{psfig}

\def \rsun {\ifmmode$R$_{\odot}\else R$_{\odot}$\fi}

\def \hcm {\hbox {\ifmmode $ H atoms cm$^{-2}\else H atoms cm$^{-2}$\fi}}
\def\approxgt{\mathrel{\hbox{\rlap{\lower.55ex \hbox {$\sim$}}
        \kern-.3em \raise.4ex \hbox{$>$}}}}
\def\approxlt{\mathrel{\hbox{\rlap{\lower.55ex \hbox {$\sim$}}
        \kern-.3em \raise.4ex \hbox{$<$}}}}
\newcommand {\rosat} {{ROSAT }}

\newcommand {\exosat} {{EXOSAT }}

\newcommand {\sax} {{\it BeppoSAX }}
\newcommand {\ginga} {{\it GINGA }}

\newcommand {\etal} {et~al. }

\newcommand {\ergcms} {erg cm$^{-2}$ s$^{-1}~$}
\newcommand {\chisq} {$\chi ^{2}$}

\begin{document}

   \thesaurus{         
              (; 
               )} 
   \title{The Complex 0.1-100 keV X-Ray Spectrum of PKS 2155-304 }


   \author{P. Giommi\inst{1}, F. Fiore\inst{1,2}, M. Guainazzi\inst{1},
M. Feroci\inst{3}, F. Frontera \inst{4,5}, G. Ghisellini\inst{6}, 
P. Grandi\inst{3}, L. Maraschi\inst{6}, T. Mineo\inst{7}, S. Molendi\inst{8,1}, 
A. Orr\inst{9}, S. Piraino\inst{7}, A. Segreto\inst{7}, 
G. Tagliaferri\inst{6}, A. Treves\inst{10}
}

   \offprints{P. Giommi}

   \institute{
   {\sax Science Data Center, ASI,
    Via Corcolle, 19
    I-00131 Roma , Italy
   }
\and
   {Osservatorio Astronomico di Roma, Monteporzio Catone,
    Italy
   }
\and
   {Istituto di Astrofisica Spaziale, CNR,
    Via Enrico Fermi, 21
    I-00044 Frascati, Italy
   }
\and
   {Dipartimento di Fisica, Universit\'a di Ferrara, Via Paradiso 12,
   I-44100 Ferrara, Italy
   }
\and
   {Istituto TESRE, CNR, Via Gobetti 101,
    I-40129 Bologna, Italy
   }
\and
   {Osservatorio Astronomico di Brera, 
    via Brera 28,
    I-20121 Milano, Italy
   }
\and
   {Istituto di Fisica Cosmica ed Applicazioni dell'Informatica, C.N.R.
    Via Ugo La Malfa, 156
    Palermo, Italy
   }
\and
   {Istituto di Fisica Cosmica, CNR, 
   Via Bassini 15
   I-20133 Milano, Italy
   }
\and
   {Space Science Department of ESA, 
    ESTEC,
    Keplerlaan,
    Noordwijk, The Netherlands
   }
\and
   {Universita' di Milano, sede di Como, 
   Via Lucini 3, I-22100 Como, Italy
   }
}
   \date{Received  ; accepted }

   \maketitle

   \markboth{Giommi et al., The Complex 0.1-100 keV X-Ray Spectrum of PKS 2155-304}{}

   \begin{abstract}

A long ($>100,000$ seconds) observation of the bright BL Lac object PKS 2155-304
has been carried out with the Narrow Field Instruments of the \sax satellite
as part of the Science Verification Phase. 
The source was detected between 0.1 and about 100
keV at an intermediate intensity level compared to previous observations. 
The unique spectral coverage of \sax has allowed us to detect a number of 
spectral features.
Between 0.1 and 10 keV the spectrum can be well described by a convex spectrum 
with (energy) slope gradually steepening from 1.1 to 1.6. 
At higher energies evidence for a sharp spectral hardening is found, while 
in the soft X-rays (0.1-1.0 keV) some evidence for an absorption feature 
was found. 
Indication for an emission line at 6.4 keV in the source 
rest frame is present. 
Repeated variability of $\approx 20-30$\% around the mean flux 
is clearly detected on time scales of a few hours. 
From the symmetry and 
timescale of the observed variations we derive limits on the magnetic field 
and on the maximum energy of the emitting particles, implying that 
PKS 2155-304 should not be bright at TeV energies.


   \end{abstract}

\section{Introduction}

BL Lac objects are highly variable AGN emitting non-thermal radiation 
from radio to high energy gamma-rays, in some cases up to TeV energies. 
The mechanism responsible for the production of radiation over such a wide 
energy range is believed to be the synchrotron emission followed at 
higher energies by inverse Compton scattering (e.g. Bregman \etal 1994). 
PKS 2155-304 is a BL Lac that was discovered in the X-ray band 
where it is one of the brightest extragalactic sources. Its synchrotron power 
output peaks at UV frequencies (Giommi, Ansari \& Micol 1995) after which 
a steep, highly variable,
continuum follows up to the mid-hard X-ray band where a much flatter 
component must emerge to explain the gamma ray emission detected 
by CGRO (Vestrand \etal 1995).

PKS 2155-304 has shown absorption features
in the soft X-rays that have been attributed to highly ionized oxygen, 
possibly moving at relativistic speed toward the observer 
(Canizares \& Kruper 1984, Madejski \etal 1991, 1997). 
Evidence for other absorption features have also been found at EUV frequencies 
by K\"onigl \etal (1995).

In this letter we report the results of a long ($>$ 2 days) 
observation of PKS 2155-305 that was carried out on 20-22 November 1996
as part of the \sax 
Science Verification Phase (SVP). Here we concentrate
on the analysis of the general shape of the energy spectrum and we briefly 
describe the time variability. A detailed timing analysis and 
a comprehensive study of the soft X-ray absorption features will be presented 
in future papers.

\section{Observations}

The \sax X-ray telescopes include one Low Energy Concentrator 
Spectrometer (LECS, Parmar \etal 1997) sensitive in the 0.1-10 keV band, and 
three identical Medium Energy Concentrator Spectrometers (MECS, Boella \etal
1997) covering the 1.5-10 keV band.
During the \sax SVP PKS 2155-304 was observed for more than 2 days using all 
the Narrow Field Instruments. The effective exposure time in the MECS was 107,702 seconds.
Since at the LECS instrument can only be operated when the spacecraft is 
in the earth shadow 
the exposure time in this instrument was limited to 36,303 seconds. 
The data analysis was based on the linearized, cleaned event files 
obtained from the \sax SDC on-line archive (Giommi \& Fiore 1997) 
and on the XIMAGE package (Giommi \etal 1991) upgraded to support the 
analysis of \sax data. 
The average count rate in the MECS was 1.66 cts/s (three units) and 1.59 cts/s 
in the LECS, after corrections for the instrument PSF and dead time.   

The Phoswich Detector System (PDS, Frontera \etal 1997) is made up of four 
units, and was operated in collimator rocking mode, with a pair of
units pointings to the source and the other pair pointing at the background,
the two pairs switching on and off source every 96 seconds.
The net source spectra have been obtained by subtracting the `off' from the 
`on' counts.
The data from the four units have been summed after gain equalization.
The Seyfert galaxy NGC7172 is located at about 1.6 degrees from
PKS2155-304, but we do not expect any detectable flux contamination in the 
1.3 degrees FWHM field of view of the PDS. 
The High Pressure Gas Scintillation Proportional Counter (HPGSPC, Manzo \etal 
1997) 
observation  resulted in an upper limit of $5\times 10^{-12}$ \ergcms 
between 10 and 20 keV, which is consistent with the PDS data in the same 
energy range.

\section{Time and spectral variability}

PKS 2155-304 is well known to be a highly variable source.
During the ten \exosat observations, which were carried out over a period 
of two years, PKS 2155-304 changed its flux by about  
a factor 10 (e.g. Giommi \etal 1990) from $\approx 2.\times 10^{-11}$ to 
$\approx 2.\times 10^{-10} $ \ergcms in the 2-10 keV band.  
The average flux seen by the \sax imaging instruments of $\approx 6.\times 
10^{-11} $ \ergcms (2-10 keV) is therefore a medium flux level.  
A detailed timing analysis of the \sax observation of PKS 2155-304
will be presented elsewhere. Here we briefly describe and comment on the main 
features seen.

\begin{figure}
\epsfig{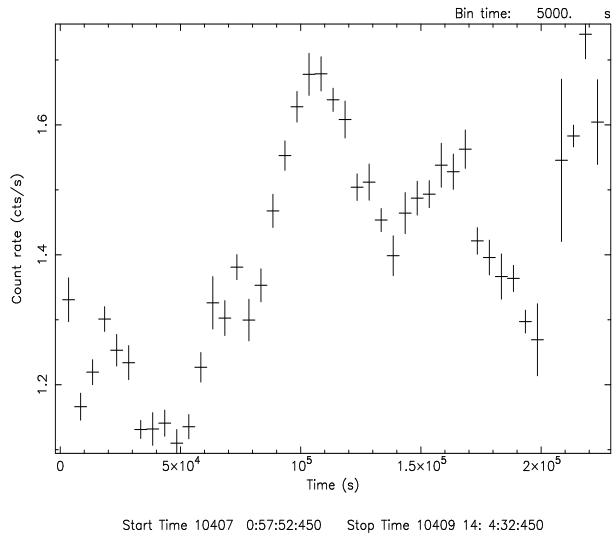}
\caption{The 1.5-10 keV MECS lightcurve of PKS2155-304. The plotted count rate 
is $\approx 90\% $ of the actual one due to the 4 arcminutes extraction 
radius for the lightcurve data}
\label{fig1}
\end{figure}

A visual inspection of the light curve of PKS 2155-304 (see figure ~\ref{fig1}) 
shows rapid (timescale of $\approx 10{^4}$ seconds) and moderate amplitude 
(up to about 50\% peak to peak) intensity variations. Single flares are 
clearly resolved with rise time approximately equal to decay time suggesting 
that the geometry within the emitting region is driving the observed light 
curve.

The PKS 2155-304 spectral slopes seen during \exosat and \ginga observations 
(Giommi \etal 1991, Sambruna \etal 1994, Sembay \etal 1993) 
are a strong function of intensity 
up to a flux of $\approx 6-8\times 10^{-11}$ \ergcms above which the spectral
slope does not change much with flux level. Spectral variability was not 
observed during \rosat observations when PKS2155-304 was 
found to be in a very high intensity state (Brinkmann \etal 1994).
When X-ray spectral variability is present the spectrum hardens
 when the source 
intensity increases, a behaviour that is typical of HBL BL Lacertae objects 
where the synchrotron break is in the UV/X-ray band (Giommi \etal 1990, 
Padovani \& Giommi 1995).
At the medium flux level seen during the \sax observation 
spectral variations not larger than $\approx 0.1$ in slope are to be 
expected for the range of variability observed (see figure 10 in 
Sembay \etal 1993). 
A simple power law fit to the \sax data confirms that the spectral slope 
changed less than 0.1 from the brightest to the lowest intensity state. 
In the following we study the 0.1-100 keV spectrum using data from the 
entire observation.

\begin{figure}
\epsfig{figure=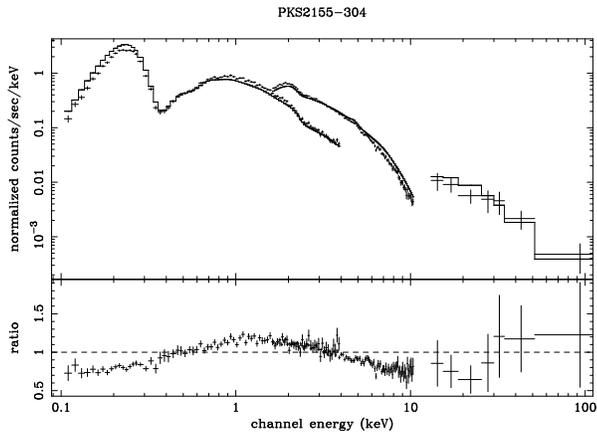,height=9.0cm, angle=-90}

\caption{Fit to a power law spectrum. The residuals from the best fit model 
clearly show spectral curvature up to 10 keV and an excess above 10-20 keV}
\label{fig2}
\end{figure}

\begin{figure}
\epsfig{figure=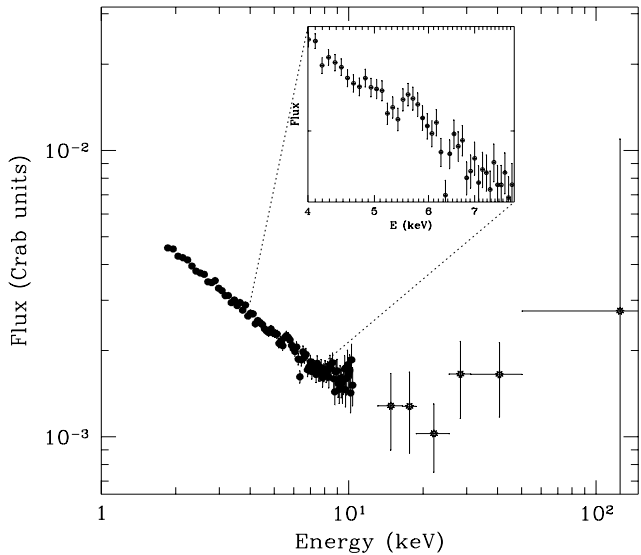,height=8.5cm, width=9.0cm}

\caption{The spectrum of PKS 2155-304 divided by the spectrum of 
the Crab Nebula.
A flux excess at around 5.7 keV (6.4 keV in the rest frame of PKS 2155-304) is 
apparent}
\label{fig3}
\end{figure}

\section{Spectral fits}

\subsection{The 1-10 keV continuum} 

Spectra were extracted from a 9.0 arcmin radius circular region around 
PKS2155-304 for the LECS, (which allows the detection of 95\% of the 
photons at 0.28 keV) and from a 4 arcmin region for the MECS which includes 
about 90\% of the flux. Spectral analysis was performed using the XSPEC 
package and the best instrument calibration available in November 1997. 
In all cases the $N_H$ has been constrained to be equal or 
higher than the Galactic value along the line of sight of $1.36\times10^{20}$
cm$^{-2}$ (measured by F.J Lockman. See Madejski \etal 1997).

Figure ~\ref{fig2} shows the 0.1-100 keV data from the LECS, MECS and PDS
instruments together with a fit to a simple power law model.
The residuals from the best fit clearly show a curvature between 0.1 and 8-10 keV  and an excess above $\approx 20 $ keV.
The best fit power law energy slope, which is a reasonable representation of 
the data only in the 1.5-10 keV band, is well within the range 
observed with \exosat (Treves \etal 1989) and close to 
the one measured with \ginga (Sembay \etal 1993) when PKS2155-304 was observed 
at the same flux level of the \sax observation.
The simple power law model over the entire band is clearly unacceptable so 
we also fitted the
data with more complex models. Table 1 summarises the best fit parameters for a 
power law, a broken power law and a ``curved spectrum'' defined as 
$$F(E)=E^{-(f(E)*\alpha_E + (1-f(E))*\alpha_H)} $$
where $f(E)=(1-exp(-E/E_0))^{\beta} $
and $\alpha_E$ and $\alpha_H$ are the low and high energy asymptotic slopes
(G. Matt, private communication). 
The \chisq values are unacceptably large for the power law and the broken 
power law model and the improvement 
obtained using the curved spectrum is highly significant.

More complex fitting involving one or more absorption features have been 
attempted. The last line in table 1 shows that the addition of one notch
to the ``curved'' model reduces the \chisq to 188.5 at the expenses of adding 
three parameters. An F test shows that the \chisq improvement is significant
at more than 99\% level. We note however, that the magnitude of the feature is
very close to the current calibration limit of the LECS instrument and more 
observations are probably needed to confirm the detection.

\subsection{The hard tail}

The power law fit to joint LECS, MECS and PDS data (see figure ~\ref{fig2}) 
shows an excess of radiation above the steep 2-10 keV power law from 
$\approx 20~keV$ and up to the highest energy where the source is detected.
A different way of showing the presence of the hard component is presented in 
figure ~\ref{fig3} where  the MECS--PDS spectrum of PKS 2155-304 is divided 
by the \sax spectrum of the Crab Nebula obtained with the same instruments. 
This method provides a direct way of visualising the shape of a spectrum 
and avoids potential problems in the cross-calibration of the instruments since 
uncertainties apply in equal way to both sources. 
Since the soft X-ray spectrum 
of PKS 2155-304 is steeper than that of the Crab Nebula, the data in 
figure ~\ref{fig3} follow a steep power law up to about 20 keV followed 
by a sharp hardening with a slope similar or even flatter than that of 
the Crab Nebula ($\alpha \simeq 1.$). A fit to the 20--200 PDS points with 
a power law yields an energy index of $0.7 \pm 0.7$.

\begin{table*}
\centering
\caption{PKS 2155-304 spectral fits}
\begin{tabular}{lcccccccc}
\hline \hline
model    & $N_H^a$  &  $\alpha_E$ & $\alpha_H$ & $E_{break}^b$ & $\beta$
& $E_{notch}^b$ &cov. frac.  
& $\chi^2$ (dof)\\
\hline
PL        &  2.6$+/-$0.1 & 1.60$+/-$0.01 & -- & -- &-- &-- &--& 774.1 (181) \\
Broken PL &  1.46+/-0.04  & 1.24$+/-$0.03 & 1.68$+/-$0.02 & 1.4$+/-$0.2 
& -- &-- &--& 237.0 (179) \\
Curved    &  1.36+0.02 & 1.09$+/-$0.02 & 1.63$+/-$0.02 & 2.3$+/-$0.3 
& 0.78 &-- &--& 194.0 (178) \\
Curved+ notch &  & & & & &0.55$+/-$0.04& 0.05$^{+0.06}_{-0.03}$ & 188.5 (175)\\
\hline
\end{tabular}
 
$^a$ in $10^{20}$ cm$^{-2}$; $^b$ in keV
\end{table*}

\subsection{A 6.4 keV Iron line? }

From figure ~\ref{fig3} we see that there is indication for 
an emission feature around 5.7 keV, consistent 
with a Fe $K_{\alpha}$ line from cold Fe in the rest frame of PKS 2155-304
(z=0.116 Falomo, Pesce \& Treves 1993). 
We tested the significance of the line detection both on the original data and 
on the spectral ratio between PKS 2155-304 and Crab. Adding of a Gaussian 
emission profile to a simple
power-law continuum in the 4.-7.5 keV band yields a $\Delta \chi^2 = 11$ for 79
d.o.f. corresponding to an F test significance of $> 99 \%$. 
Rest frame line parameters are $E = 6.40^{+0.12}_{-0.13} \ keV$ and $EW =
40^{+30}_{-20} \ eV$. Although the statistical significance is rather high
and no significant calibration problems are known around 5.7 keV, we feel 
that more systematic analysis of \sax data is necessary for a final 
confirmation of the $K_{\alpha}$ line.

\section{Discussion}

The analysis of the average broad-band X-ray spectrum of PKS 2155-304 in 
a medium intensity state shows a sharp break above 10-20 keV, and a 
possible detection of a Fe $K_{\alpha} $ line. 
Intensity variability is limited to $\approx \pm 25\%$ around a mean value 
and does not affect the presence of the hard tail.
Statistical evidence for a low energy absorption feature was  
also found close to the calibration limit.  

A hard tail above 10 keV was reported  in 
PKS 2155-304 during one HEAO1 observation (Urry \& Mushotsky 1982) but 
\ginga observations did not confirm such a finding (Sembay \etal 1993).
A spectral hardening in the hard X-ray band is expected from the SSC mechanism 
and is necessary to explain the gamma ray emission detected by CGRO (Vestrand 
\etal 1995).

The detection of a Fe $K_{\alpha}$ line indicates the presence of cold gas,
possibly associated with an accretion disk (George \& Fabian 1991). 
Disk emission from PKS 2155-304 might have become detectable in the X-ray band 
because this part of the spectrum is no longer strongly dominated by the 
synchrotron component which peaks at UV frequencies (see however the 
constraints posed by simultaneous optical UV photometry, Urry et al 1993).
In this scenario the hard tail could include a component due to Compton 
reflection from the cold disk. 

From the light curve it can be seen that the flares have approximately equal
rising and decaying timescales, with no indications of a plateau.
It is highly unlikely that the two timescales correspond
to the particle acceleration and cooling timescales, since
a priori there is no reason for them to be equal. 
Instead, these timescales may be associated with the dimension of
the emitting region, and hence with the light crossing time, provided that 
both the electron injection and cooling processes operate on time--scales 
shorter than $R/c$ (where $R$ is the source radius).
In this scenario the light crossing time is the only characteristic
time that can be observed (Massaro \etal, 1996, Ghisellini \etal 1997).
Assuming that the flares are due to quasi--instantaneous
injection of electrons and rapid cooling,
we require that the synchrotron and self Compton cooling times
are equal to or shorter than $R/c$:
$
t_{cool}\, = \, { 6\pi mc^2 \over \sigma_Tc\gamma_o B^2(1+U_S/U_B)}
\, \le { R \over c},
$
where $\sigma_T$ is the Thomson scattering cross section, $\gamma_o m_ec^2$
is the energy of the particles emitting at the observed frequency, 
$U_B=B^2/(8\pi)$
is the magnetic energy density, and $U_S$ is the radiation energy density
of the synchrotron emission (as measured in the comoving frame).
The ratio $U_S/U_B$ measures the relative importance of the self--Compton
to the synchrotron luminosity, and in PKS 2155--304 it
can be estimated by making the ratio of the $\gamma$--ray luminosity
(Vestrand \etal 1995) 
to the optical--UV luminosity, where the synchrotron spectrum peaks.
The energy $\gamma_0$ is related to the observed frequency
$\nu_o \simeq 3.7\times 10^6 \gamma_o^2 B \delta/(1+z)$
(assuming synchrotron radiation), and a limit on $R$ is found using 
$R\sim ct_{var}\delta/(1+z)$, where $\delta$ is the beaming factor.
We then obtain the two limits:
$
B\, \ge  \, 1.16 \, \nu_{17}^{-1/3} t_h^{-2/3}(1+U_S/U_B)^{-2/3}
\left({1+z \over \delta}\right)^{1/3} ,
$
$
\gamma_o\, \le \, 1.5\times 10^5 \nu_{17}^{2/3}
\left[{ t_h (1+U_S/U_B)(1+z) \over \delta }\right]^{1/3} ,
$
where $B$ is in Gauss, $t_h$ is the variability timescale
measured in hours, and $\nu_o=10^{17}\nu_{17}$ Hz.
Assuming $\delta=10$, $t_h=8$, $\nu_{17}=1$ and $U_S/U_B=0.5$,
we find $B\ge 0.11$ G and $\gamma_o\le 1.6\times 10^5$.
The upper limit on the energy is particularly important, since
it immediately corresponds to an upper limit on the maximum photon
energies that can be emitted (by, e.g. the inverse Compton process): 
$h\nu_{obs} < \gamma_om_ec^2 \delta/(1+z) = 182 \, \delta^{2/3}$ GeV.
We then conclude that PKS 2155--304 should not be an important 
TeV emitter, unlike Mkn 421 and Mkn 501, unless a very high degree
of beaming is present.

\begin{acknowledgements}

We thank G. Fossati and F. Haart for their contribution to 
the calibration of the low energy response of the LECS instrument. We 
are also grateful to G. Matt for providing the code used for the
``curved spectrum'' model.

\end{acknowledgements}


\begin{thebibliography}{}

   \bibitem{} Boella G. \etal 1997 A\&AS, 122, 327 
   \bibitem{} Bregman J. 1994, in Multi-Wavelength Continuum Emission of AGN,
T.J.L. Courvoisier and A. Blecha eds., IAU Symp no. 159, p. 5
   \bibitem{} Brinkmann W. \etal 1994, A\&A, 288, 433
   \bibitem{} Canizares, C.R., Kruper J. 1984 ApJ, 278, L99 
   \bibitem{} Falomo, R., Pesce, J. E., \& Treves, A. 1993, APJ 411, L63
   \bibitem{} Frontera F. \etal 1997 A\&AS, 122, 357
   \bibitem{} K\"onigl A., Kartje J.F., Bowyer S., Kahn, S.M., and Hwang C.Y.
1995, ApJ 446, 598
   \bibitem{} George, I.M., \& Fabian, A. C. 1991, MNRAS, 249, 352
   \bibitem{} Ghisellini, G., Villata, M., Raiteri, C.M. \etal 1997, submitted 
to A\&A
   \bibitem{} Giommi P., Barr P., Garilli B., Maccagni D., Pollock A. 1990, ApJ,
356, 455
   \bibitem{} Giommi, P., Angelini, L., Jacobs, P. \& . Tagliaferri G.  1991
          in "Astronomical Data Analysis Software and Systems I",
          D.M.Worrall, C. Biemesderfer and J. Barnes eds,
          1991, A.S.P. Conf. Ser. 25, 100 
   \bibitem{} Giommi P., Ansari, S.G., \& Micol, A. 1995, A\&AS 109, 267
   \bibitem{} Giommi P.,\& Fiore F. 1997, Proc. of "5th International Workshop 
on Data Analysis in Astronomy, Erice.
   \bibitem{} Madejski G.M., Mushotzky R.F., Weaver K.A., Arnaud K.A. 1991, ApJ 
370, 198
   \bibitem{} Madejski G.M. \etal 1997 ApJ, submitted.
   \bibitem{} Manzo G. \etal 1997 A\&AS, 122, 341
   \bibitem{} Massaro, E., Nesci, R. Trevese, D. \etal, 1996, A\& A 314, 87
   \bibitem{} Padovani, P., Giommi, P. 1995 ApJ, 444, 567
   \bibitem{} Parmar A. \etal 1997 A\&AS, 122, 309
   \bibitem{} Stark A. A., Gammie C. F., Wilson R. W., Bally J., Linke R. A.,
 Heiles C., Hurwitz M. 1992 ApJS, 77 
   \bibitem{} Sambruna R.M., Barr P., Giommi P., Maraschi L., Tagliaferri,
        A. Treves 1994 ApJ, 434,468
   \bibitem{} Sembay, S, \etal 1993, ApJ 404,112
   \bibitem{} Treves A. \etal 1989, ApJ, 341, 733
   \bibitem{} Urry, C.M., \& Moshotsky R.F. 1982, ApJ, 253, 38
   \bibitem{} Urry, C.M. \etal 1993, ApJ, 411, 614
   \bibitem{} Vestrand, W.T., Stacy, G., Sreekumar, P. 1995, ApJ 454, L93
 
\end{thebibliography}
\end{document}